\documentstyle[12pt]{article}
\begin{document}

\begin{center}

{\large \bf An analysis of cosmological perturbations\\[5pt]
in hydrodynamical and field representations}\\

\vspace{2cm}

J\'ulio C. Fabris$^{a} $
\footnote{e-mail:\quad fabris@cce.ufes.br},
Sergio V.B. Gon\c{c}alves$^{b}$
\footnote{e-mail:\quad svbg@if.uff.br}
and Nazira A. Tomimura$^{b}$
\footnote{e-mail:\quad nazt@if.uff.br}
\\
\vspace{0.6cm}
{\it $^{a}$Departamento de F\'{\i}sica, Universidade Federal do
Esp\'{\i}rito Santo,} \\
\vspace{0.1cm}
{\it Vit\'oria CEP 29060-900-Esp\'{\i}rito Santo. Brazil.} \\
\vspace{0.4cm}
{\it $^{b}$Instituto de F\'{\i}sica, Universidade Federal
Fluminense}\\
\vspace{0.1cm}
{\it Niter\'oi CEP 24210-340-Rio de Janeiro. Brazil}
\vspace{1cm}

\end{center}

\centerline{\bf abstract}
Density fluctuations of fluids with negative pressure exhibit decreasing
time behaviour in the long wavelength
limit, but are strongly unstable in the small wavelength limit when a
hydrodynamical
approach is used. On the other hand, the corresponding gravitational waves
are
well behaved. We verify that the instabilities present in density fluctuations are due
essentially to
the hydrodynamical representation; if we turn to a field
representation
that lead to the same background behaviour, the instabilities are no more
present. In the long wavelength limit, both approachs give the same results.
We show also that this inequivalence between background and perturbative
level is
a feature of negative pressure fluid. When the fluid has positive
pressure,
the hydrodynamical representation leads to the same behaviour as the
field
representation both at the background and perturbative levels.

\vspace{1.0cm}
\par
PACS number: 98.80.Bp, 95.35.+d
\par
keywords: cosmology, perturbations theory, large scale structure.
\vspace{2cm}

\section{Introduction}

Fluids with negative pressures have become of utmost importance in modern
cosmology.
They were first considered in the context of the inflationary scenario in
which the
early universe has a very short period of accelerating expansion
\cite{guth}. This inflationary phase
solves many problems of the standard model which are connected to specific
choices of
initial conditions, like the flatness and horizon problem. At the same
time, the origin
of the seeds of the large scale structures observed today in the universe
has a natural
explanation in the inflationary scenario, that considers them as quantum
fluctuations
of a scalar field in a de Sitter background. The inflationary phase must
end up with
a transition to the radiative phase of the standard model. In many aspects, the
inflationary
scenario is plagued with problems of arbitrary choice of parameters of the
microphysics,
but its great success with the above mentioned problems led the community
to consider
it as part of the standard scenario.
\par
More recently, the observation of the supernova of high redshift led to
the preliminary
conclusion that the Universe today is in an accelerating expansion
\cite{perlmutter,riess}.
This is a 
surprising result since there was little
doubts that the Universe was desacelerating and that 
the desaceleration parameter $q$ was positive.
If the results of the supernova observations are confirmed, the energy of
the Universe
today must be dominated by a fluid with negative pressure. It could be a
cosmological
constant, but other cases are not excluded, like a fluid of cosmic string
or domain walls
\cite{kamio,davis,stel}
or some scalar field, with a peculiar kind of potential, called
quintessence
\cite{ioav,cald,wang,peebles,jerome}.
\par
In many pratical cases, these fluids with negative pressure may be
represented by a perfect
fluid with a barotropic equation of state $p = \alpha\rho$, with $\alpha <
0$. To be more
precise, in order to have an accelerating universe the strong energy
condition must
be violated and $\alpha < - \frac{1}{3}$. A fluid of cosmic string leads
to $\alpha = - \frac{1}{3}$, representing the limiting case between an
accelerating and desaccelerating
universe; $\alpha = - \frac{2}{3}$ and $\alpha = - 1$ represent respectively a fluid of
domain wall and a
cosmological constant. The quintessence fluid only approximativelly can be
expressed
by a barotropic fluid with $\alpha < 0$.
\par
In \cite{julio} perturbations in fluids with negative pressure were
studied in the
hydrodynamical representation. It was found that when the strong energy
condition
is violated, there are instabilities in the small wavelength limit. This
result can
be understood by remembering that in this limit the expansion of the
universe plays
no important role: the negative pressure at this limit acts in the same way as gravity and nothing can prevent
the collapse.
This can be
easily seen
in the case of a cosmic string fluid for which the density contrast behaves
as
$\Delta \propto t^{1 \pm \sqrt{1 + \frac{n^2}{3}}}$. When $n
\rightarrow \infty$
divergences appear, leading to the instability of
the background model. In the cosmological perturbation theory, we are generally interested
on the unstable modes, but that are not divergent. From here on, instabilities will refer
to these undesirable divergent modes.
\par
This study of density perturbations for a fluid of cosmic string has been
extended
in \cite{ser}, where a two-fluid model was considered, one of them being
the string
fluid and the other ordinary matter with positive pressure. Special
attention was payed to
the case of a closed
spatial section. From the point of view of the background, such models have
many
interesting features, mainly conected to the horizon problem and to the
age of
the universe. However, the study of scalar fluctuations around this background shows
the presence of instability in the small scale limit, as in the case of the
one fluid model. On the other hand, the
study
of gravitational waves for these models \cite{ser1} reveals a
very
regular behaviour because gravitational waves are mainly connected with
the scale factor behaviour, being quite insensitive to the matter
representation. This indicates that the instabilities detected in \cite{julio,ser} could
be due to the
hydrodynamical representation, which could disappear in a more fundamental
approach.
\par
In fact, the hydrodynamic representation for fluids with negative pressure is frequently a
very poor approximation. Negative pressures appear in situations involving
phase transitions in a primordial universe (topological defects) or a fundamental self-interacting
scalar field. The exact formulation involves consequently fields, and a representation
using a fluid with a barotropic equation of state only in very special situations
may be employed. The employment of a perfect fluid representation, mainly when
fluids of negative pressure are involved,
can be viewed as a practical simplification which in many situations
gives the same results as those that could be obtained by employing a
more fundamental field representation.
\par
The main point is that the equivalence of a hydrodynamical representation
with a more fundamental one is very restricted and it can lead to
complete misleading results depending on the problem treated. The instability
in the small wavelength limit quoted above is an example. Fluids with negative
pressure should have a field representation where their
main features could be retained. The representation we will investigate here
involves the more reasonable coupling of gravity with a scalar field
with a self interacting term.
\par
The aim of this paper is to show that
the scalar field representation can more conveniently retain the
features we could expect from fluids with negative pressure, mainly for those that
are interesting for cosmology, as the objects resulting from phase transitions,
like cosmic string.
We will concern mainly with the consequences
of these
different representations for an analysis of perturbations around a homogenous and
isotropic
expanding universe. We verify that when we use a field
representation,
with the same behaviour for the scale factor as the corresponding
hydrodynamical
representation, the instabilities present in the latter case are absent in
the former one.
This is due essentialy to the fact that when we pass from a hydrodynamical
representation
to a field representation of the fluid we also pass from an Euler's type
equation to
a Klein-Gordon's type equation, and there is no correspondence when the
pressure is negative. On the other hand, for large scale perturbations, both
approachs lead to the same results.
\par
We begin by analysing a stiff matter fluid which can 
mimic a scalar field in a very simple way. As it is well known, a free scalar
field minimally
coupled to gravity reproduces the stiff matter equation of state. We will
show in section $2$
that for the stiff matter and free scalar field models, the agreement
between the results
occurs not only at the background level but also at the perturbative level.
In section $3$ we extend this analysis to a perfect fluid with
arbitrary equation of state $p = \alpha\rho$: we determine the
corresponding field representation, showing that, at perturbative level,
the equivelence between these two approachs remains only when
$\alpha > 0$.
In section $4$ we review briefly the results obtained in \cite{ser,ser1}
and we discuss the possibility of a field representation for this two fluid
model.
In section $5$, we review a model of variable cosmological constant that
leads, from the point of view of the evolution of the scalar factor, to
the same behaviour
as a cosmic string fluid. In section $6$, we perform a perturbative
analysis of the
variable constant model, and show explicitly that they are free of
instabilities,
both for scalar perturbations and tensor perturbations.
In section $7$, we discuss the results.

\section{Free scalar field model}

The most simple gravity model with a scalar field is described by the 
action 
\begin{equation}
\label{a}
{\cal{S}}=\int \sqrt{-g}[R - \phi,_{\alpha}\phi,^{\alpha}]d^4 x \quad.
\end{equation}
It represents a free scalar field minimally coupled to
gravity.
The field equations obtained in accordance with the principle of least
action by varying $S$ with respect to the dynamical variables are:
\begin{equation}
\label{b}
R_{\mu\nu} - \frac{1}{2}g_{\mu\nu}R = \phi_{;\mu}\phi_{;\nu} -
\frac{1}{2}g_{\mu\nu}\phi_{;\rho}\phi^{;\rho}\quad,
\end{equation}
\begin{equation}
\label{c}
\Box\phi = 0\quad.
\end{equation}
In a Friedmann-Robertson-Walker flat space-time (FRW), the metric
describing the
four dimensional geometry is
\begin{equation}
ds^2 = dt^2 - a^2(t)\biggr(dx^2 + dy^2 + dz^2\biggl) \quad ,
\end{equation}
and the field equations take the form
\begin{eqnarray}
\label{d}
3\biggr(\frac{\dot a}{a}\biggl)^2 &=& \frac{1}{2}\dot\phi^{2}\quad,\\
\label{e}
2\frac{\ddot a}{a} + \biggr(\frac{\dot a}{a}\biggl)^2 &=& -
\frac{\dot\phi^{2}}{2} \quad,\\
\label{f}
\ddot\phi + 3\frac{\dot a}{a}\dot\phi &=& 0 \quad ,
\end{eqnarray}
where the overdot denotes the derivative with respect to the time
coordinate $t$. The equations
(\ref{d},\ref{e},\ref{f}) are not independent due to the Bianchi
identities.
\par
On the other hand, in an universe filled with a perfect fluid
we have the
field equations
\begin{eqnarray}
\label{g}
R_{\mu\nu} - \frac{1}{2}g_{\mu\nu}R &=& 8\pi GT_{\mu\nu} \quad ,\\
\label{g1}
T^{\mu\nu}_{;\mu} &=& 0 \quad ,
\end{eqnarray}
where
\begin{equation}
\label{h}
T_{\mu\nu} = (\rho + p)u_\mu u_\nu - pg_{\mu\nu} \quad ,
\end{equation}
with $p = \alpha\rho$, $\alpha$ being a constant. The most common values of
$\alpha$ of cosmological interest
are $0$ (pressurelless matter), $\frac{1}{3}$ (radiation) and $1$ (stiff
matter). Negative values
of $\alpha$ has acquired increasing importance due to the inflationary
paradigm and the new results coming
from the supernova type Ia observations, as it was discussed before.
Topological defects also require a negative equation of state.
Again, equations (\ref{g},\ref{g1}) are connected by the Bianchi identities.
\par
In the same FRW flat background, (\ref{g},\ref{g1}) take the form
\begin{eqnarray}
\label{d1}
3\biggr(\frac{\dot a}{a}\biggl)^2 &=& 8\pi G\rho\quad,\\
\label{e1}
2\frac{\ddot a}{a} + \biggr(\frac{\dot a}{a}\biggl)^2 &=& - 8\pi Gp \quad,
\\
\label{f1}
\dot\rho + 3(1 + \alpha)\frac{\dot a}{a}\rho &=& 0 \quad.
\end{eqnarray}
\par
It is a straightforward to check that the equations
(\ref{d},\ref{e},\ref{f}) and equations (\ref{d1},\ref{e1},\ref{f1})
permit the identification
\begin{equation}
\label{i}
\rho_{\phi} = p_{\phi}= \frac{{\dot\phi}^2}{2},
\end{equation}
i.e., the "scalar field fluid" behaves like a stiff matter in the
hydrodynamical approach. The scalar field sound velocity, in this case, is
$c_{\phi}^{2}=\dot p_{\phi}/\dot\rho_{\phi}=1$.
The scale factor both in the free scalar field and hydrodynamical stiff
matter cases behaves
as $a \propto t^{1/3}$.
\par
The evaluation of the perturbed quantities follows
the well known approach of Lifshitz and Khalatnikov
\cite{lifshitz,weinberg}. It can be treated either with the synchronous
gauge or the gauge-invariant formalism. Here we choose to work in the
synchronous gauge formalism, where $h_{\mu 0} = 0$.
\par
We study first density perturbations and then gravitational waves for this
free field model.

\subsection{Density perturbations}

Introducing small perturbations around the background solutions, the
perturbed equations for the scalar-tensor model read
\begin{equation}
\label{l}
\ddot h + 2\frac{\dot a}{a}\dot h = 4\dot\phi\dot{\delta\phi}\quad,
\end{equation}
\begin{equation}
\label{m}
\ddot{\delta\phi} + 3\frac{\dot a}{a}\dot{\delta\phi} +
\frac{n^{2}}{a^{2}}\delta\phi - \frac{1}{2}\dot h\dot\phi = 0
\quad ,\\
\end{equation}
where $h = h_{kk}/a^{2}$ and $n^{2}$ comes from the Helmholtz equation
$\nabla^2{\cal Q} + n^{2}{\cal Q} = 0$: the scalar functions ${\cal
Q}(x^{k})$ are the eingefunctions of the three-dimensional Laplacian
operator.
\par
In order, to solve equations (\ref{l},\ref{m}) it is more convenient to work
in the conformal time,
$dt = ad\eta$. The scale factor behaves as
$a \propto \eta^{1/2}$. In terms of this new time parameter the solution
of the perturbed equations is given by
\cite{tossa}
\begin{equation}
\label{n}
\delta\phi = \eta^{-3/2}\int\eta^{3/2}\biggr(c_1(n)J_{1}(n\eta) +
c_2(n)N_1(n\eta)\biggl)d\eta \quad,
\end{equation}
where $J_{1}$ and $N_{1}$ are, respectively, the Bessel and Neumann
functions of the first order, and $c_1(n)$, $c_2(n)$ are two arbitrary
constants.
\par
We need to verify if the evolution of the perturbations are specified by
the equation $\delta p_{\phi} = \alpha\delta\rho_{\phi}$ as 
it happens with
the background evolution showed previously. This should imply that the adiabatic
approximation is verified here. To do this, we consider the
perturbation of the equation (\ref{b}) and (\ref{h})
\begin{displaymath}
\delta G^{\mu\nu} = \delta\phi,^{\mu}\phi,^{\nu} + 
\phi,^{\mu}\delta\phi,^{\nu} +
\frac{1}{2}h^{\mu\nu}g^{\alpha\beta}\phi,_{\beta}
\phi,_{\alpha}
\end{displaymath}
\begin{equation}
\label{dd}
+ \frac{1}{2}g^{\mu\nu}h^{\alpha\beta}\phi,_{\beta}
\phi,_{\alpha} - \frac{1}{2}g^{\mu\nu}g^{\alpha\beta}
(\delta\phi,_{\beta}\phi,_{\alpha} + \phi,_{\beta}
\delta\phi,_{\alpha})\quad,
\end{equation}
\begin{equation}
\label{ee}
\delta T^{\mu\nu} = (\delta\rho_{\phi} + \delta p_{\phi})
U^{\mu}U^{\nu} + (\rho_{\phi} + p _{\phi})(\delta U^{\mu}U^{\nu} +
U^{\mu}\delta U^{\nu}) - \delta p_{\phi}g^{\mu\nu} +
p_{\phi}h^{\mu\nu}\quad .
\end{equation}
Using the synchronous gauge condition, we have
\begin{equation}
\delta G_{00} = \dot\phi\delta\dot\phi\quad,
\end{equation}
\begin{equation}
\delta G_{ij} = -\frac{1}{2}h_{ij}\dot\phi^{2} +
a^{2}\delta_{ij}\dot\phi\delta\dot\phi\quad,
\end{equation}
\begin{equation}
\delta T_{00} = \delta\rho_{\phi}\quad,
\end{equation}
\begin{equation}
\delta T_{ij} = -p_{\phi}h_{ij} + a^{2}\delta_{ij}\delta p_{\phi}\quad.
\end{equation}
If we consider the perturbed Einstein equations $\delta G_{\mu\nu} = 8\pi
G\delta T_{\mu\nu}$ and the values of
$\rho_\phi$ and $p_{\phi}$ obtained from equations (\ref{d1}) and (\ref{e1}),
we have that
\begin{equation}
\delta\rho_{\phi} = \delta p_{\phi}\quad.
\end{equation}
\par
In the hydrodynamical approach, the solution for the Einstein perturbed
equations with $p = \rho$ and $\delta p = \delta\rho$, leads to the expression
\cite{julio}
\begin{equation}
\label{sm}
\Delta = \frac{\delta\rho}{\rho} =
\eta^{-3/2}\int\eta^{5/2}\biggr(d_1(n)J_0(n\eta) +
d_2(n)N_0(n\eta)\biggl)d\eta \quad ,
\end{equation}
where $d_1(n)$ and $d_2(n)$ are the integration constants.
Remembering that $\rho_\phi = \frac{{\phi'}^2}{2a^2}$ and $\delta\rho_\phi
= 2\frac{\phi'\delta\phi'}{a^2}$,
the quantity $\Delta_\phi = \frac{\delta\phi'}{\phi'}$, computed from
(\ref{n}) reduces to (\ref{sm}), using
simple recurrence relations for Bessel's functions.
\par
Hence, in this simple model where the matter is described by a scalar
field, the background and perturbed equations can be put in a barotropic
form. The "velocity of sound" $\delta p_{\phi}/\delta\rho_{\phi}$ is the
same as the one defined by $c_{\phi}^{2}=\dot p_{\phi}/\dot\rho_{\phi}$ in agreement
with \cite{grishchuk}, where it is shown that this result corresponds to the
low-frequency regime of the scalar field oscillations.

\subsection{Gravitational waves}

Here, the perturbed equation of gravitational waves is:
\begin{equation}
\label{pipi}
h'' - 2\frac{a'}{a}h' + \biggl[n^{2} - 2\frac{a''}{a} +
2\frac{a'^{2}}{a^{2}}\biggr]h = 0\quad,
\end{equation}
where $h_{ij} = h(\eta)Q_{ij}$, $Q_{ij}$ being a tracelless transverse
eigenfunction on the
spatial section, and primes denote derivative with respect to conformal
time $d\eta = adt$.
\par
After inserting the backgroung solutions, we obtain the solution of the
equation (\ref{pipi}), as follow:
\begin{equation}
\label{papa}
h = \eta\biggl(e_1(n)J_1(n\eta) + e_2N_1(n\eta)\biggr)\quad,
\end{equation}
where $e_1(n)$ and $e_2(n)$  are the integration constants. This solution is
valid for
both representations.
\par
It is easy to verify that the above solution is well-behaved and stable.

\section{Field representation for an arbitrary barotropic equation of state}

Let us return to the Einstein's equations coupled to a perfect fluid,
with a barotropic equation of state $p = \alpha\rho$. Solving the Einstein's
equation for a flat spatial section, we obtain for
the scale factor $a = a_0t^\frac{2}{3(1 + \alpha)}$.
Let us now consider a self interacting scalar field coupled to gravity.
The Lagrangian has the form,
\begin{equation}
L = \sqrt{-g}\biggr[R - \phi_{;\rho}\phi^{;\rho} + 2V(\phi)\biggl]
\end{equation}
where the potential $V(\phi)$ represents the self interaction term.
The field equations are
\begin{equation}
R_{\mu\nu} - \frac{1}{2}g_{\mu\nu}R = \phi_{;\mu}\phi_{;\nu} - \frac{1}{2}g_{\mu\nu}\phi_{;\rho}\phi^{;\rho} +
g_{\mu\nu}V(\phi) \quad , \quad \Box\phi = - V_\phi(\phi) \quad,
\end{equation}
where $V_\phi$ means partial derivative of the potential with respect to $\phi$.
\par
This scalar-tensor model may lead to the same behaviour of the scale factor as in the perfect fluid model
provided that the potential takes the form,
\begin{equation}
V(\phi) = \frac{2}{3}\frac{(1-\alpha)}{(1 + \alpha)^{2}}\exp({\mp\sqrt{3(1+\alpha)}\phi}) \quad.
\end{equation}
Indeed, for a FRW background, this potential leads to the solutions
\begin{equation}
a = a_0t^\frac{2}{3(1 + \alpha)} \quad , \quad \phi = \pm \frac{2}{\sqrt{3(1 + \alpha)}}\ln t \quad .
\end{equation}
For the case of a cosmic string fluid, $a \propto t$, the potential is $V(\phi) = - 2e^{-\sqrt{2}\phi}$.
\par
We now turn to the perturbative level.
The perturbation of the scalar-tensor model leads to
\begin{eqnarray}
\label{p1}
h'' + \frac{a'}{a}h' &=& 4\phi'\delta\phi' + 2\biggr(\phi'' + 2\frac{a'}{a}\phi'\biggl)\delta\phi \quad , \\
\label{p2}
\delta\phi'' + 2\frac{a'}{a}\delta' + \biggr(n^2 + V_{\phi\phi}a^2\biggl)\delta\phi &=& \frac{1}{2}\phi'h' \quad ,
\end{eqnarray}
where we follow the same definitions as before and we have employed the conformal time.
\par
Combining equations (\ref{p1},\ref{p2}), and inserting the background solutions, we obtain the third order
differential equation
\begin{eqnarray}
\delta\phi''' + \frac{7 + 3\alpha}{1 + 3\alpha}\frac{\delta\phi''}{\eta} +
\biggr\{n^2 + \biggr[\frac{2 - 24\alpha - 18\alpha^2}{(1 + 3\alpha)^2\eta^2}\biggl]\biggl\}\delta\phi'\nonumber\\
+ \biggr\{3\frac{1 + \alpha}{1 + 3\alpha}\frac{n^2}{\eta} - 18\frac{1 - \alpha^2}{(1 + 3\alpha)^2}\frac{1}{\eta^3}
\biggl\}\delta\phi = 0 \quad .
\end{eqnarray}
One solution for this equation is $\delta\phi = \eta^{-3\frac{1+\alpha}{1+3\alpha}}$, which is given
by the residual coordinate transformation freedom characteristic of the synchronous coordinate condition.
Using this known solution and some suitable transformation, we can reduce the third order equation to
a second order Bessel equation. The final solution for $\delta\phi$ is
\begin{equation}
\delta\phi = \eta^{-3\frac{1+\alpha}{1+3\alpha}}\int c_\pm\eta^\frac{3}{2}J_{\pm\nu}(n\eta)d\eta \quad ,
\end{equation}
where $c_\pm$ are integration constants, that in general depend on $n$, and $J_{\pm\nu}$ is a Bessel function of order
$\nu = \frac{5 + 3\alpha}{2(1 + 3\alpha)}$.
\par
The solution for the density contrast in the case of the hydrodynamical representation, with $p = \alpha\rho$ is well
known\cite{julio}. It can be written as
\begin{eqnarray}
\label{s1}
\Delta &=& \eta^{-3\frac{1+\alpha}{1+3\alpha}}
\int \eta^\frac{5}{2}\biggr(c_+J_{\mu}(\sqrt{\alpha}n\eta) +
c_-J_{-\mu}(\sqrt{\alpha}n\eta)\biggl)d\eta \quad ,
\alpha > 0 \quad , \\
\label{s2}
\Delta &=& \eta^{-3\frac{1+\alpha}{1+3\alpha}}\int \eta^\frac{5}{2}\biggr(c_+I_{\mu}(\sqrt{-\alpha}n\eta) +
c_-K_\mu(\sqrt{-\alpha}n\eta)\biggl)d\eta \quad , \quad \alpha < 0 \quad ,
\end{eqnarray}
where now $I_\mu$ and $K_\mu$ are the modified Bessel functions, and $\mu = \frac{3}{2}\biggl(\frac{1 - \alpha}{1 + \alpha}\biggr)$.
In order to connect both results we must remember that $\Delta_\phi = \frac{\delta\rho_\phi}{\rho_\phi}$,
where $\rho_\phi = \frac{{\dot\phi}^2}{2} + V(\phi)$. Using some Bessel's recurrence relations, as in the
previous section we find
\begin{equation}
\label{s3}
\Delta_\phi = C_\pm\eta^{-3\frac{1+\alpha}{1+3\alpha}}\biggr\{(1 - \frac{1}{\alpha})\eta^\frac{5}{2}J_{\pm\mu}(n\eta)
+ \int \eta^\frac{5}{2}J_{\pm\mu}(n\eta)d\eta\biggl\} \quad .
\end{equation}
This expression differs from (\ref{s1}) by the factor $\alpha$ in the argument of the Bessel function, and by the
first term in (\ref{s3}). However, when
$\alpha > 0$ solutions (\ref{s1},\ref{s3}) have the same behaviour in the large and small
wavelength limit. For $\alpha = 1$, the equivalence between the two approaches is
complete, as can be easily checked by comparing (\ref{s3}) with (\ref{s1}) for this special case.
For $\alpha < 0$, the correspondence between solutions (\ref{s2},\ref{s3}) occurs only for $n \rightarrow 0$ (large scale perturbations);
for $n \rightarrow \infty$ (small scale perturbations) the hydrodynamical representation exhibits strong
instabilities while the scalar field representation displays accoustic oscillation.
\par
Finally we remark that both representations give the same behaviour for gravitational waves.

\section{Perturbations in cosmic string fluid}

An important particular case of the scalar-tensor model develloped in the previous section concerns
the case of cosmic string.
A cosmic string fluid may mimic a perfect fluid with an equation of
state
$p = - \frac{\rho}{3}$. In \cite{julio} the fate of density perturbation
in fluids
with negative pressure has been studied. The main conclusion was that,
in the long wavelength approximation, there are only decreasing modes when
the equation of state is such that the strong energy condition is
violated;
for small wavelengths, instabilities can arise due to the fact that 
the pressure
contributes in the same sense as gravity. For the equation of state of
a cosmic string fluid displayed above, density perturbations behave as
\begin{equation}
\Delta = t^{-1 \pm \sqrt{1 + \frac{n^{2}}{3}}} \quad .
\end{equation}
Hence, in the small wavelength limit, $n \rightarrow \infty$,
instabilities
come to scene.
\par
A scalar-tensor representation of the cosmic string fluid is
given by a scalar tensor model with
$V(\phi) = 2\exp(\pm\sqrt{2}\phi)$. A perturbation analysis as it was performed in section $3$ leads to
the following expression for the perturbation in the scalar field (the integration procedure follows very
closely to that sketched in the previous section):
\begin{equation}
\Delta_\phi \propto t^{-1 \pm \sqrt{1 - n^2}} \quad .
\end{equation}
In the large wavelength limit both representations give the same results; in the
small wavelength limit the hydrodynamical representation display instabilities while the
scalar-tensor model exhibits accoustic oscillations.
\par
In \cite{ser} it was considered the case of two non interacting  fluid,
one of them represented by an equation of state like $p = -(1/3)\rho$
while
the
second fluid is the ordinary matter with a
barotropic equation of state $p = \alpha\rho$. This may represent in more
realistic way the presence of fluids with negative pressure in the universe.
The equations of motion are
\begin{equation}
3(\frac{\dot a}{a})^{2} + 3\frac{k}{a^{2}} = 8\pi G(\rho_{m} +
\rho_{s})\quad,
\end{equation}
\begin{equation}
2\frac{\ddot a}{a} + (\frac{\dot a}{a})^{2} + \frac{k}{a^{2}} = \frac{8\pi
G}{3}(\rho_{s} - 3\alpha\rho_{m})\quad,
\end{equation}
\begin{equation} 
\dot\rho_{m} + 3\frac{\dot a}{a}(1 + \alpha)\rho_{m} = 0\quad,
\end{equation}
\begin{equation}
\dot\rho_{s} + 2\frac{\dot a}{a}\rho_{s} = 0\quad.
\end{equation}
\par
In this equations $\rho_m$ and $\rho_s$ denote the ordinary fluid and
stringlike fluid
densities respectivelly.
\par
The background solutions are:
\begin{equation}
\label{caca}
a = a_0\sinh^2(\frac{\sqrt{\gamma}}{2}\eta)\quad,\mbox{when}\quad\alpha = 0\quad;
\end{equation}
\begin{equation}
\label{cece}
a = a_0\sinh(\sqrt{\gamma}\eta)\quad,\mbox{when}\quad\alpha = 1/3\quad;
\end{equation}
\begin{equation}      
\label{cici}
a = a_0\sqrt{\sinh(2\sqrt{\gamma}\eta)}\quad,\mbox{when}\quad\alpha = 1\quad,
\end{equation}
where $\gamma = \vert\frac{8}{3}\pi G\rho_{0s} - k\vert$, $\rho_{0s}$ is defined as the ratio
$\rho_{s} = \frac{\rho_{0s}}{a^2}$, $\eta$ being the conformal time.

\subsection{Density perturbations}

Perturbing the Einstein's equations and imposing the syncrhonous
coordinate condition, a set of equations for the ordinary matter and the
cosmic fluid perturbations is obtained:
\begin{equation}
\ddot h + 2\frac{\dot a}{a}\dot h = -6\frac{\ddot a}{a}\Delta_{m}\quad,
\end{equation}
\begin{equation}
\dot\Delta_{m} + (1 + \alpha)(\Psi - \frac{\dot h}{2}) = 0\quad,
\end{equation}
\begin{equation}
(1 + \alpha)\dot\Psi + (1 + \alpha)(2 - 3\alpha)\frac{\dot a}{a}\Psi -
\frac{n^{2}}{a^{2}}\alpha\Delta_{m} = 0\quad,
\end{equation}
\begin{equation} 
\dot\Delta_{s} + \frac{2}{3}(\Theta - \frac{\dot h}{2}) = 0\quad,
\end{equation}
\begin{equation}
\dot\Theta + 3\frac{\dot a}{a}\Theta + \frac{n^{2}}{2a^{2}}\Delta_{s} =
0\quad.
\end{equation}
\par
The integration procedure is standard \cite{ser}, so we will just present
the final results:
\begin{center}
\begin{tabular}{|c||c|c|c|}
\hline
model  &  perturbations  &  material phase $\quad (t\rightarrow 0)$  &
\quad string phase$(t\rightarrow\infty)\quad$                  \\
\hline\hline

$\alpha = 0$  
                 &  $\Delta_{m}$ &  $\eta^{2}$  &
$\mbox{constant}$  
                 \\       
                 \cline{2-4}
                 &  $\Delta_{s}$  &
$\frac{1}{\eta^{\frac{2}{3}}}(a_{1}
                     K_{\frac{3}{2}}(n\eta) +
a_{2}I_{\frac{3}{2}}(n\eta))$  &
                    $t^{-1\pm\sqrt{1 +
\frac{n^{2}}{3}}}$                                   \\ \hline
$\alpha = 1/3$
                 &  $\Delta_{m}$ &
$\frac{1}{\eta^{\frac{1}{2}}}
                     \int\eta^{\frac{5}{2}}\biggr[c_{1}
                     J_{\frac{1}{2}}(n\eta) + c_{2}
                     J_{-\frac{1}{2}}(n\eta)\biggl]
                     d\eta$  &  $\cos(\frac{n}{\sqrt{3}}
                     \ln t)$  
                   \\       
                   \cline{2-4}
                 &  $\Delta_{s}$  &
$\frac{1}{\eta^{\frac{1}{2}}}(d_{1}
                     I_{\frac{1}{2}}(n\eta) +
d_{2}K_{\frac{1}{2}}(n\eta))$  &
                    $t^{-1\pm\sqrt{1 +
\frac{n^{2}}{3}}}$                                    \\ \hline
$\alpha = 1$      
                 &  $\Delta_{m}$ &  
                    $\frac{1}{\eta^{\frac{3}{2}}}
                     \int\eta^{\frac{5}{2}}\biggr[c_{1}
                     J_{0}(n\eta) + c_{2}
                     J_{0}(n\eta)\biggl]
                     d\eta$  &
$\cos(n\ln t)$  
                  \\       
                  \cline{2-4}
                 &  $\Delta_{s}$  &
$d_{1}I_{0}(n\eta) +
d_{2}K_{0}(n\eta))$  &
                    $t^{-1\pm\sqrt{1 +
\frac{n^{2}}{3}}}$                                   \\ \hline   
\end{tabular}
\end{center}
\par
We must exhibit the behaviour of the
perturbative model for $t\rightarrow 0$ and $t\rightarrow\infty$ to find
an instability in the density perturbations of cosmic string fluid. Indeed, for 
small wavelengths, i.e. $n\rightarrow\infty$, the modified Bessel functions $I_{\nu}(x)$
and $K_{\nu}(x)$ go as $e^{\pm x}/\sqrt{x}$. However, the ordinary
matter has a very regular behaviour for $t\rightarrow 0$ and
$t\rightarrow\infty$.
\par
The instabilities presented above can be attributed to the matter content
of the universe and,
mainly, to the approach used to describe the density perturbations. We
will see that these instabilities do not appear in gravitational waves
because the matter content of the universe plays a less important role,
being only sensitive to the behaviour of scale factor.
\par
In fact, the same behaviour for the scale factor described above can be obtained by a scalar-tensor
model, with a suitable potential, coupled to ordinary matter.
For example, in the case of pressurelles ordinary matter ($p = 0$), the results for the background
can be recovered if the potential reads
\begin{equation}
V(\phi) = V_0\sinh^{-4}(C\phi) \quad , \quad V_0 = \frac{8\Omega^2 + 2}{{a_0}^2} \quad , \quad C = \frac{1}{\sqrt{8\Omega^2 + 2}}
\quad , \quad \Omega = \frac{2\pi G\rho_0}{3} \quad,
\end{equation}
where $\rho_0$ is defined by the first integral of the conservation of the
energy-moment tensor for
the pressurelless fluid, $\rho = \frac{\rho_0}{a^3}$.
We note that the scalar field representation for the two fluid model (pressurelles matter plus
string fluid) requires a different potential
with respect to the case of the one fluid model (only string fluid).
\par
At the perturbative level, the effect of replacing the barotropic fluid with negative pressure
by a scalar-field model
is quite the same as that described in section $3$. With respect with the preceding table
of solutions, the modified Bessel functions must be replaced by ordinary Bessel functions for
the perturbation in the exotic fluid.
Hence the instability present in the small wavelength limit disappear again;
the behaviour in the long wavelength limit is not changed.

\subsection{Gravitational waves}

Following the work \cite{ser1}, we have
the perturbed equations that govern the evolution of gravitational waves
\begin{equation}
\label{popo}
h'' - 2\frac{a'}{a}h' + \biggl[\tilde n^{2} - 2\biggl(\frac{a''}{a} -
\frac{a^{2}}{a^{2}}\biggr)\biggr] = 0\quad,
\end{equation}
where $\tilde n^{2} = (1/r^{2})(n^{2} + 2k)$ and prime here denotes derivative
with respect to $\theta = r\eta$.
\par
After inserting the background solutions (\ref{caca},\ref{cece},\ref{cici}),
equation (\ref{popo}) can be generally rewritten in terms of a
hypergeometric equation. Its final solutions for different phases of the
evolution of the universe read as follows:
\begin{center}
\begin{tabular}{|c||c|c|}
\hline
model  &  $\quad h_{1}\quad$  &  $\quad h_{2}\quad$                  \\
\hline\hline

$\alpha = -1$  
                 &  $\sqrt{x^{2}-1}\biggl[\frac{1 +
x}{2}\biggr]^{-2+\sqrt{1-\tilde n^{2}}}\times$
                 &  $\sqrt{x^{2}-1}\biggl[\frac{1 +
x}{2}\biggr]^{-2-\sqrt{1-\tilde n^{2}}}\times$
                 \\       
                 &  ${_{2}F_{1}}\biggl(2-\sqrt{1-\tilde
n^{2}},\frac{1}{2}-\sqrt{1-\tilde n^{2}},$
                 &  ${_{2}F_{1}}\biggl(\frac{1}{2}+
\sqrt{1-\tilde n^{2}},2+\sqrt{1-\tilde n^{2}},$
                 \\
                 &  $1-2\sqrt{1-\tilde n^{2}},
\frac{2}{1+x}\biggr)$  
                 &  $1+2\sqrt{1-\tilde n^{2}},
\frac{2}{1+x}\biggr)$  
                 \\       
                 \hline
$\alpha = 1/3$
                 &  $\exp((\sqrt{1-\tilde  n^{2}})\eta)\sinh\eta$   
                 &  $\exp(-(\sqrt{1-\tilde  n^{2}})\eta)\sinh\eta$                      
                 \\
                 \hline
$\alpha = 0$
                 &  $\sqrt{x^{2}-1}\biggl[\frac{1 +
x}{2}\biggr]^{1+\sqrt{4-\tilde n^{2}}}\times$
                 &  $\sqrt{x^{2}-1}\biggl[\frac{1 +
x}{2}\biggr]^{1-\sqrt{4-\tilde n^{2}}}\times$
                 \\
                 &  ${_{2}F_{1}}\biggl(-1-\sqrt{4-
\tilde n^{2}},\frac{1}{2}-\sqrt{4-\tilde n^{2}},$
                 &  ${_{2}F_{1}}\biggl(\frac{1}{2}
+\sqrt{4-\tilde n^{2}},-1+\sqrt{4-\tilde n^{2}},$
                 \\
                 &  $1-2\sqrt{4-\tilde n^{2}},
\frac{2}{1+x}\biggr)$  
                 &  $1+2\sqrt{4-\tilde n^{2}}, \frac{2}{1+x}\biggr)$
\\
                 \hline
$\alpha = 1$      
                 &  $\sqrt{x^{2}-1}\biggl[\frac{1 +
x}{2}\biggr]^{(-1+\sqrt{1-4\tilde n^{2}})/2}\times$
                 &  $\sqrt{x^{2}-1}\biggl[\frac{1 +
x}{2}\biggr]^{(-1-\sqrt{1-4\tilde n^{2}})/2}\times$
                 \\
                 &  ${_{2}F_{1}}\biggl(\frac{1-
\sqrt{1-4\tilde n^{2}}}{2},\frac{1-\sqrt{1-4\tilde  n^{2}}}{2},$
                 &  $\times\quad_{2}F_{1}\biggl(\frac{1+\sqrt{1-
4\tilde n^{2}}}{2},\frac{1+\sqrt{1-4\tilde n^{2}}}{2},$
                 \\
                 &  $1-\sqrt{1-4\tilde n^{2}},
\frac{2}{1+x}\biggr)$                  
                 &  $1+\sqrt{1-4\tilde n^{2}},
\frac{2}{1+x}\biggr)$                  
                 \\  \hline   
\end{tabular}
\end{center}
where in the above expressions, ${_{2}F_{1}}$ is a hypergeometric function
and $x = \cos (r\theta)$, $r$ being
a constant.
\par
In this two-fluid model, the behaviour of gravitational waves in a closed
universe exhibits, in what concerns the behaviour of
the scale factor, the dynamic of an open universe. It would cause,
also, distorsion in the spectrum of the CMBR. The determination of the
evolution of perturbations and its connection with this distorsion is
technically difficult to be evaluated. But, for the proposal of the present
work, the fundamental feature to be retained in the above solutions is
that
they do not exhibit instabilities.

\section{Time dependent cosmological constant model}

We present for completeness the background equations based on the reference\cite{rivera} which is the traditional
scalar-tensor theory with a
potential that is equivalent to a time dependent cosmological constant model.
This is one of the cases which can be represented phenomenologically by
a fluid with an equation of state $p = - \frac{\rho}{3}$ in what concerns the
behaviour of the scale factor. The action is
given by
\begin{equation}
\label{o}
{\cal S} = \frac{1}{16\pi G}\int \sqrt{-g}[\phi R - \phi^{-1}\omega
g^{\mu\nu}\partial_{\mu}\partial_{\nu}\phi + 2\phi\Lambda(\phi)]d^{4}x +
{\cal S}_{ng}\quad.
\end{equation}
We remark, however, that in the present case we have a non-minimally coupled scalar field,
in opposition to the models described before.
\par
The field equations are
\begin{equation}
\label{p}
G_{\mu\nu} = \frac{8\pi T_{\mu\nu}}{\phi} + \Lambda(\phi)g_{\mu\nu} +
\omega\phi^{-2}(\phi,_{\mu}\phi,_{\nu} -
\frac{1}{2}g_{\mu\nu}\phi,_{\alpha}\phi,^{\alpha}) +
\phi^{-1}(\phi;_{\mu\nu} - g_{\mu\nu}\Box\phi)\quad,
\end{equation}
\begin{equation}
\label{q}
\Box\phi + \frac{2\phi^{2}d\Lambda/d\phi - 2\phi\Lambda(\phi)}{3 +
2\omega} = \frac{1}{3 + 2\omega}\biggr(8\pi T -
\frac{d\omega}{d\phi}\phi,_{\alpha}\phi,^{\alpha}\biggl)\quad.
\end{equation}
\par
We shall consider the case where $\omega =$ constant, $\Lambda(\phi) =
c\phi^{m}$ and $\phi = \phi_{1}t^{q}$, with $c$, $m$ and $q$ constants.
The ansatz allow us to obtain analytical solutions in the form of
power law, which leads
to closed expressions for the perturbative equations.
\par
The energy-momentum tensor describes an ordinary perfect fluid, such that
\begin{equation}
\label{noe}
T^{\mu\nu};_{\nu} = 0\quad.
\end{equation}
The equations of motion are
\begin{equation}
\label{r}
3\frac{\dot a^{2}}{a^{2}} + 3\frac{k}{a^{2}} - c\phi^{m} = 
\frac{8\pi\rho}{\phi} + \frac{\omega}{2}\frac{\dot\phi^{2}}{\phi^{2}} -
3\frac{\dot a}{a}\frac{\dot\phi}{\phi}\quad,
\end{equation}
\begin{equation}
\label{s}
-2\frac{\ddot a}{a} - \frac{\dot a^{2}}{a^{2}} - \frac{k}{a^{2}} +
c\phi^{m} = \frac{8\pi p}{\phi} +
\frac{\omega}{2}\frac{\dot\phi^{2}}{\phi^{2}} + \frac{\ddot\phi}{\phi} +
2\frac{\dot a}{a}\frac{\dot\phi}{\phi}\quad,
\end{equation}
\begin{equation}
\label{t}
\frac{\ddot\phi}{\phi} + 3\frac{\dot a}{a}\frac{\dot\phi}{\phi} =
\frac{2c(1-m)\phi^{m}}{3 + 2\omega} + \frac{8\pi (\rho - 3p)}{\phi (3 +
2\omega)}\quad.
\end{equation}
\\

The background solutions are

\begin{center}
\begin{tabular}{|c||c|c|c|c|c|}
\hline
model  &  curvature  &  $\quad a(t)\quad$  &  $\phi(t)$  &  $\rho$  &
$\quad\Lambda(t)\quad$ \\ \hline\hline

$\rho = 0$  
                 &  $k\ne 0$ &  $a_{1}t$  
                 &  $\phi_{1}t^{\frac{-2}{m}}$
                 &  -  &  $\Lambda_{1}/t^{2}$  \\
\cline{2-6}
                 &  $k = 0$  &  $a_{1}t$                                 &
$\phi_{1}t^{\frac{-2}
                   {1\pm\sqrt{3 + 2\omega}}}$ 
                 &  -  &  $\Lambda_{1}/t^{2}$ \\ \hline
$p = \alpha\rho$
                 &  any $k$  &  $a_{1}t$  
                 &  $\phi_{1}t^{-(1 + 3\alpha)}$  
                 &  $s/a^{3(1 + \alpha)}$  
                 &  $\Lambda_{1}/t^{2}$ \\ \hline
$p = -\rho$      
                 &  any $k$  &  $a_{1}t$ 
                 &  $\phi_{1}t^{2}$  &  $\rho_{0}
= \mbox{const.}$  
                 &  $\Lambda_{1}/t^{2}$ \\ \hline  
\end{tabular}
\end{center}
where $a_{1}$, $\phi_{1}$, $\Lambda_{1}$, and $s$ are constants.
\par
The solutions for the scale factor above are characteristic of an equation of state
$p_{\phi} = -\frac{1}{3}\rho_{\phi}\quad$.
\par
We remark that in all above solutions the scale-factor behaves as $a
\propto t$
corresponding to a "coasting" universe.
From the point of view of the background behaviour, the above
solutions exhibit the same characteristics as the perfect fluid
formulation.

\section{Perturbations in a time dependent cosmological constant model}

We extend our perturbative analysis for this model, first for density
perturbations and then for gravitational waves.
 
\subsection{Density perturbations}

The perturbed equations for the time dependent cosmological constant model
are:
\begin{displaymath}
\ddot h + 2\frac{\dot a}{a}\dot h = \frac{16\pi}{\phi}(\Delta
-\lambda)\biggr(\frac{3\alpha\omega+3\alpha+\omega+2}{3+2\omega}\biggl)\rho
\end{displaymath}
\begin{equation}
\label{x}
- 2cm\phi^{m}\biggr(\frac{m + 2\omega + 2}{3 + 2\omega}\biggl)\lambda +
2\ddot\lambda + 4\frac{\dot\phi}{\phi}(1 + \omega)\dot\lambda\quad,
\end{equation}
\begin{displaymath}
\ddot\lambda +\biggr(2\frac{\dot\phi}{\phi} + 3\frac{\dot
a}{a}\biggl)\dot\lambda + \biggr(\frac{\ddot\phi}{\phi} + 3\frac{\dot
a}{a}\frac{\dot\phi}{\phi}\biggl)\lambda -
\frac{1}{2}\frac{\dot\phi}{\phi}\dot h
\end{displaymath}
\begin{equation}
\label{y}
+ \frac{n^{2}}{a^{2}}\lambda + \frac{2c(m^{2}-1)\phi^{m}}{3 +
2\omega}\lambda = \frac{8\pi}{(3 + 2\omega)\phi}\Delta(1 -
3\alpha)\rho\quad,
\end{equation}
\begin{equation}
\label{z}
\dot\Delta = (1 + \alpha)\biggr(\frac{1}{2}\dot h - \delta
U^{k},_{k}\biggl)\quad,
\end{equation}
\begin{equation}
\label{aa}
\frac{\partial}{\partial t}\biggr(a^{5}\delta U^{k}(1 + \alpha)\rho\biggl)
= -a^{3}\alpha\rho\Delta,^{k}\quad,
\end{equation}
where $h = h_{kk}a^{-2}$, $\lambda = \delta\phi/\phi$ and $\Delta =
\delta\rho/\rho$. All functions are spatially expanded into spherical
harmonics $f(x,t) = f(t){\cal Q}$, with $\nabla^2{\cal Q} =
-n^{2}{\cal Q}$.
\par
Next, we solve the above equations for the vacuum fluid ($\alpha = -1$),
the empty universe ($\rho = 0$), radiation phase ($\alpha = 1/3$), and
dust phase ($\alpha = 0$). We use the background relations in order to
simplify the final system of equations.

\subsubsection{Vacuum fluid phase ($\alpha = -1$)}

In this case $\Delta = 0$ and we have the following equations:
\begin{equation}
\label{cc}
\ddot h + \frac{2}{t}\dot h = (1 + 2\omega)\frac{\lambda}{t^2} +  8(1 +
\omega)\frac{\dot\lambda}{t} + 2\ddot\lambda\quad,
\end{equation}
\begin{equation}
\label{ccc}
\ddot\lambda + 7\frac{\dot\lambda}{t} + \biggl[8 + \frac{n^{2}}{a_{1}}\biggl]\frac{\lambda}{t^2} -
\frac{1}{t}\dot h = 0\quad,
\end{equation}
with the solution
\begin{equation}
\label{cccc}
\lambda = t^{p}\quad \mbox{and}\quad h\propto t^{p}\quad,\quad
\mbox{where}\quad p = -2\pm\sqrt{8\omega -\frac{n^{2}}{a_{1}^{2}}}\quad.
\end{equation}
There is no divergent behaviour in the small wavelength limit.

\subsubsection{Empty universe ($\rho = 0$)}

Here, the system of second order differential equations is
\begin{equation}
\label{bb}
\ddot h + \frac{2}{t}\dot h = - \frac{4}{m}\biggr[\frac{m + 2\omega + 2}{3 + 2\omega}\biggl]\frac{\lambda}{t^2}
- \frac{8(1 +
\omega)}{m}\frac{\dot\lambda}{t}+ 2\ddot\lambda\quad,
\end{equation}
\begin{equation}
\label{bbb}
\ddot\lambda + \biggl[\frac{3m - 4}{m}\biggr]\dot\lambda +
\biggl[\frac{n^2}{{a_1}^2} - 4(2 + m)\frac{1 - m}{m^2}\biggl]\frac{\lambda}{t^2} + \frac{1}{m}\frac{\dot h}{t} = 0\quad,
\end{equation}
whose solution is
\begin{equation}
\label{bbbb}
\lambda = t^{p}\quad\mbox{and}\quad h\propto t^{p}\quad
\mbox{where}\quad p = \frac{1}{m}\biggl(1 - m \pm\sqrt{9 + 6m - 3m^2 + 8\omega - m^{2}\frac{n^{2}}{a_{1}^{2}}}\biggr)\quad,
\end{equation}
where $m$ is arbitrary for $k \neq 0$ and $m = 1 \pm \sqrt{3 + 2\omega}$ for $k = 0$.
As in the preceding case, the solutions are stable.

\subsubsection{Radiation phase ($\alpha = 1/3$)}

Combining
conveniently the
perturbed equations, we find a fifth order Euler's type equation for
$\lambda$:
\begin{eqnarray}
\lambda^v + 11\frac{\lambda^{iv}}{t} +
\biggr[\frac{4}{3}\frac{n^2}{{a_1}^2} + 31 - 2d -
2\frac{k}{a_1^{2}}\biggl]\frac{\stackrel{...}{\lambda}}{t^2}
+ \biggr[\frac{16}{3}\frac{n^2}{{a_1}^2} + 22 - 8d -
8\frac{k}{a_1^{2}}\biggl]\frac{\ddot\lambda}{t^3}
\nonumber\\
+ \biggr[\frac{1}{3}\frac{n^4}{{a_1}^4} +
\frac{n^2}{{a_1}^2}\biggr(\frac{8}{3} + \frac{2}{3}d - 2\frac{k}{a_1^{2}}\biggl)
+ 2 - 4d - 4\frac{k}{a_1^{2}}\biggl]\frac{\dot\lambda}{t^4} +
\biggr[\frac{1}{3}\frac{n^4}{{a_1}^4} +
\frac{n^2}{{a_1}^2}\biggr(\frac{2}{3}d - 2\frac{k}{a_1^{2}}\biggl)\biggl]
\frac{\lambda}{t^5} = 0 \quad ,
\end{eqnarray}
with $d = 3 + 2\omega$. Due to the residual coordinate transformation
freedom characteristic of the
synchronous coordinate condition, $\lambda \propto t^{-1}$ is a solution
of
this equation.
Hence, we can reduce it to a fourth order equation.
Supposing
a solution of the type $\lambda \propto t^r$, we find  the
polynomial equation for
$r$
\begin{equation}
r^4 + \biggr(\frac{4}{3}\frac{n^2}{{a_1}^2} - 2d - 2\frac{k}{a_1^{2}}\biggl)r^2 +
\biggr(\frac{1}{3}\frac{n^4}{{a_1}^4} - \frac{n^2}{{a_1}^2}(- \frac{2}{3}d +
2\frac{k}{a_1^{2}})\biggl) = 0
\end{equation}
whose solutions are
\begin{equation}
{r_{\pm}}^2 = - \frac{2}{3}\frac{n^2}{{a_1}^2} + d + \frac{k}{a_1^{2}} \pm \sqrt{\frac{n^4}{9{a_1}^4} + (\frac{2}{3}\frac{k}{a_1^{2}}
 - 2 d)\frac{n^2}{{a_1}^2} + (d + \frac{k}{a_1^{2}})^2} \quad .
\end{equation}
In the small wavelength limit, these solutions display an oscillatory
behaviour,
hence stability. In the longwavelength limit, 
this expression reduces to $r_\pm = \pm \sqrt{2(d + \frac{k}{a_1^{2}})}$.

\subsubsection{Dust Phase ($\alpha = 0$)}

In this case, the equations governing the evolution of density
perturbations are
\begin{eqnarray}
\ddot\Delta + 2\frac{\dot\Delta}{t} - (2 + \omega)R\frac{\Delta}{t^2} &=&
\ddot\lambda - 2(1 + \omega)\frac{\dot\lambda}{t} - (2 + \omega)S\frac{\lambda}{t^2}
\quad , \\
\ddot\lambda + \frac{\dot\lambda}{t} + \biggr[\frac{n^2}{{a_1}^2} +
S\biggl]\frac{\lambda}{t^2} &=& - \frac{\dot\Delta}{t} +
R\frac{\Delta}{t^2} \quad ,
\end{eqnarray}
where $R = \frac{2\frac{k}{{a_1}^2} - \omega - 1}{3 + 2\omega}$ and $S =
\frac{6\frac{k}{{a_1}^2} + 3 + \omega}{3 + 2\omega}$.
These equations may be solved supposing $\Delta = \Delta_0 t^r$ and
$\lambda = \lambda_0 t^r$, $r$ obeying the
polynomial equation
\begin{eqnarray}
r^4 + 2r^3 + \biggl(\frac{n^2}{{a_1}^2} + S - (3 + \omega)R - (3 +
2\omega)\biggl)r^2 \nonumber \\
+ \biggr(\frac{n^2}{{a_1}^2} - (1 + \omega)S + (3 + 2\omega)R\biggl)r
- (2 + \omega)R\frac{n^2}{{a_1}^2} = 0 \quad ,
\end{eqnarray}
This polynomial equation has no simple closed form solution. But,
numerical integration reveals the stability of the
model in the small wavelength limit.
For example, fixing $a_1 = 1$, choosing $\omega = 1$, $k = 0$ and $n = 1$ we find the roots
${r_1}^\pm \sim - 1.16 \pm 2.77i$, ${r_2}^\pm \sim 0.16 \pm 0.33i$,
while for $n = 100$, keeping unchanged the other parameters, we find
${r_1}^\pm \sim - -0.50 \pm 100i$, ${r_2}^\pm \sim -0.50 \pm 0.97i$. 

\subsection{Gravitational waves}

In this case, the perturbed equation that govern the gravitational waves
is
\begin{eqnarray}
\ddot h + \biggr(\frac{\dot\phi}{\phi} - \frac{\dot a}{a}\biggl)\dot h +
\biggr[\frac{n^2}{a^2} + 4\frac{\dot a^2}{a^2} +
2\biggr(\frac{\omega(\alpha - 1) - 1}{3 + 2\omega}
\biggl)\biggr(3\frac{\dot a^2}{a^2} + 3\frac{k}{a^2} -
\frac{\omega}{2}\frac{\dot\phi^2}{\phi^2}
+ 3\frac{\dot a}{a}\frac{\dot\phi}{\phi}\biggl)\nonumber \\
- 2 \biggr(\frac{1 + m + \omega(1 + \alpha)}{3 +
2\omega}\biggl)c\phi^m\biggl]h = 0 \quad.
\end{eqnarray}
\par
The solutions of the above equation are
\begin{eqnarray}
\rho = 0 &\rightarrow& h = C_\pm t^{\frac{1}{2}(1 + A_1 \pm \sqrt{(1 + A)^2 - 4B - 4\frac{n^2}{{a_1}^2}}}
\quad , \nonumber\\
A_1 &=& \frac{2 + m}{m} \quad , \quad B = \frac{4}{m^2}(m^2 - m - 2 - 2\omega) \quad ,\\
\alpha = - 1 &\rightarrow& h = C_\pm t^{\pm\sqrt{-(\frac{n^2}{{a_1}^2} + A_2)}} \quad , \nonumber\\
A_2 &=& \frac{1}{3 + 2\omega}\biggr[3 - 8\omega  + 4\omega^2 - 3\frac{k}{{a_1}^2}(1 + 2\omega)\biggl] \quad ,\\
\label{momo}
\alpha = 0 &\rightarrow& h = C_{\pm}t^\frac{3 \pm \sqrt{9 - 4A_3 -
4\frac{n^2}{{a_1}^2}}}{2}\quad,\quad A_3 = 2 - \frac{4k}{{a_1}^2}\\
\label{mimi}
\alpha = \frac{1}{3} &\rightarrow& h = C_{\pm}t^{2 \pm \sqrt{4\frac{k}{{a_1}^2}
- \frac{n^2}{{a_1}^2}}}\quad,
\end{eqnarray}
where $C_\pm$ are integration constants. 
\par
Here, the solution for the gravitational waves are also well-behaved and
stable.

\subsection{The General Relativity limit}

In general, the Brans-Dicke theory reduces to the General Relativity theory when
$\omega \rightarrow \infty$, except in some special cases, for example, when
the trace of the momentum-energy tensor is zero \cite{romero,faraoni}. The solutions described above
do not have a well-behaved limit when $\omega \rightarrow \infty$. In fact,
an inspection of the background equations shows that all solutions become
trivial in that limit. We could expect that in this case,
the Brans-Dicke field could become constant and the model reduces itself to General Relativity with
a cosmological constant. But, the imposition that $a \propto t$ breaks this equivalence.
\par
Concerning the perturbed solutions, they become divergent when $\omega \rightarrow \infty$.
This only express the fact that the background scenarios have no sense in this limit.
For finite $\omega$, the perturbed solutions exhibit growing and decreasing modes as usual, for
scalar and tensorial perturbations. One important feature of the solutions found before is that
when $n \rightarrow 0$, all dependence of the solutions on the wavelength of the perturbations
is carried out by the integration constants, which must be fitted correctly in order to reproduce
the power spectrum observed today.

\section{Conclusions}

In spite of the fact that fluids of negative pressure have become crucial
to understand
many theoretical and observational aspects of modern cosmology, they are
plagued with
instabilities in the small wavelength limit. These instabilities appear
mainly when
the barotropic equation of state is such that the strong energy condition
is violated.
In this paper we have exploited the possibility that these instabilities
are due
to the hydrodynamical representation. This possibility was suggested by
the fact that,
while density perturbations are bad behaved in the small wavelength limit,
gravitational
waves are well behaved in the same limit. Since gravitational waves
depend closely on
the behaviour of the scale factor but are quite insensitive to the matter
representation,
the instabilities should be due to the fact that in the hydrodynamical
representation,
the equation of state is fixed for all wavelength, while in a more
fundamental approach
it could depend on the scale of the perturbation.
\par
As a matter of fact, this possibility was first suggested in
\cite{peebles}, and it has
been studied under certains conditions in \cite{grishchuk}. In reference
\cite{peebles},
this problem has been briefly treated in the realm of minimal
scalar-tensor model
which was intended to cope with the dark matter problem through the
introduction
of a scalar field with a convenient potential. However, if we are interested in a
field that can drive the accelerated
expansion of the Universe, as it is the case in this work,
the energy of the scalar
field should
have a smooth distribution, since it should not be present in the local
clusters.
Such smooth distribution can be obtained considering that the pressure
associated to
this field is negative, such that it does not clumpsy in large scale; but
in order to avoid
instabilities at small scales, the effective equation of state should
become positive
in this limit, and small fluctuations in this field should oscillate like
an
accoustic wave.
\par
In the present work, we have verified that, when a fluid of positive
pressure can 
mimic a scalar-tensor model,
both formulations are essentially similar in the background and perturbative level.
However, when we consider a fluid of negative pressure, the equivalence
exists only
at the background level: at perturbative level, the models behave in a
complete different
way. In particular, there is agreement between both representations 
only in the large wavelength limit: for small wavelengths,
the hydrodynamical model is unstable, while the corresponding scalar field representation
exhibits accoustic oscillations. Hence, in situations where negative pressures are concerned,
a field representation leads to a much more complete scenario, being closer to a realistic model.
\par
This can be understood by remembering that when we pass from a
hydrodynamical representation
to a fundamental one based for example on scalar fields,
we change a relativistic Euler's type equation to a Klein-Gordon
equation: the
sign of the laplacian operator in these equations are the same only when
the hydrodynamical
pressure is positive; otherwise, it changes sign and, in the perfect fluid model, accoustic modes
give place to
exponential modes, resulting in the presence of instabilities.
In the long wavelength limit the laplacian operator can be neglected and the results
agree in both representation.
\par
We must notice, however, that the scalar-tensor model that corresponds to a
given perfect fluid results is quite model dependent. For example,
the one fluid model gives a potential that is different from that of a
two-fluid models.
It would be interesting to employ in a two fluid model, which is closer to a realistic
situation, the potential obtained in the one fluid model, in spite of the great
complexity of the equations. We can speculate if the resulting effective equation
of state evolves in a quite similar way as the usual quintessence models.
\par
We have extended this study for the case where the cosmic string fluid mimic
a variable cosmological constant model, derived from a non-minimal coupled scalar
field with a potential. The conclusions are basically the same as before. But, we
must stress that, in this case, there is a Brans-Dicke type coupling parameter $\omega$ that
does not recover the known General Relativity solutions when $\omega \rightarrow \infty$.
Moreover, the perturbative behaviour may become unstable in this limit, for any scale.
But, in general, for finite values of $\omega$ the perturbations do not exhibit either
the instabilities that are present in the corresponding hydrodynamical model.
\par
Finally, the fact that both approaches give the same behaviour for the long wavelength limit
even if the pressure is negative, implies that the predictions for the power spectrum of
the anisotropy of the CMBR for small values of the multipolar expansion parameter $l$, that is, for very large structures,
is not spoiled by the employment of the hydrodynamical representation.
However, for large values of $l$, that is, small angular separation, the employment of
a field representation, mainly when negative pressures are involved, seems crucial.

\section*{Acknowledgements}
We thank Nelson Pinto Neto for his suggestions and for many enlightfull discussions.
We acknowledge also the financial support from CNPq and CAPES (Brazil).


\begin{thebibliography}{50}
\bibitem{guth} S. Blau and A.H. Guth, in {\bf 300 years of gravitation},
edited
by S. Hawking and W. Israel, Cambridge University Press, Cambrdige (1987);
\bibitem{perlmutter} S. Perlmutter et al, Nature {\bf 391}, 51(1998);
\bibitem{riess} A.G. Riess et al, {\it High-z supernovae search team},
astro-ph/9805201;
\bibitem{kamio} M. Kamionkowski and N. Toumbas , Phys. Rev. Lett.{\bf 77},
587(1999);
\bibitem{davis}  R.L. Davis, Phys. Rev. {\bf D35}, 3705(1987);
\bibitem{stel} P.J. Stelmach, {\it Horizon problem in a closed Universe
dominated by fluid with negative pressure}, astro-ph/9810448;
\bibitem{ioav}  I. Waga and  A.P.M.R. Miceli, Phys. Rev. {\bf D59},
103507(1999), astro-ph/9811460;
\bibitem{cald} R.R. Caldwell, R. Dave and P.J. Steinhardt, Phys. Rev.
Lett. {\bf 80}, 1582(1998);
\bibitem{wang} I. Zlatev , I. Wang and  P.J. Stelmach,  Phys. Rev.
Lett.{\bf 82}, 896(1999), astro-ph/9807002;
\bibitem{peebles} B. Ratra and P.J.E. Peebles, Phys. Rev. {\bf D37},
3406(1988);
\bibitem{jerome} P. Brax and J. Martin, {\it Quintessence and
supergravity}, astro-ph/99050040;
\bibitem{julio} J.C. Fabris and J. Martin, Phys. Rev. {\bf D55},
5205(1997);
\bibitem{ser}J.C. Fabris and S.V.B. Gon\c{c}alves, {\it Phys. Rev.} {\bf
D56}, 6128(1997);
\bibitem{ser1}J.C. Fabris and S.V.B. Gon\c{c}alves, {\it
Mon.Not.R.Astron.Soc.} {\bf 306}, 679-683(1999);
\bibitem{lifshitz} E.M.Lifschitz and I. Khalatnikov, {\it Adv. Phys.} {\bf
12}, 185(1963);
\bibitem{weinberg} S. Weinberg, {\bf Cosmology and Gravitation},
Wiley, New York(1972);
\bibitem{tossa} J.C. Fabris and J. Tossa, Grav\&Cosm. {\bf 3}, 165(1997);
\bibitem{grishchuk} L.P. Grishchuk, Phys. Rev. {\bf D50}, 7154(1993);
\bibitem{rivera} L.O.Pimentel and L.M. Diaz-Rivera, Int. J. Mod. Phys.
{\bf  A14}, 1523(1999), gr-qc/9807016;
\bibitem{romero} C. Romero and A. Barros, Phys. Lett. {\bf A137}, 243(1993);
\bibitem{faraoni} V. Faraoni, Phys. Lett. {\bf A245}, 26(1998).
\end{thebibliography}
\end{document}